\documentclass[final]{IEEEtran} 

\usepackage{xspace}
\usepackage{pbox}
\usepackage{tabularx}
\usepackage{booktabs}
\usepackage{pdfpages}
\usepackage{graphicx}
\usepackage{times}
\usepackage{eurosym}
\usepackage[indentfirst=false,leftmargin=10pt,rightmargin=10pt]{quoting}
\usepackage[noadjust]{cite}
\renewcommand{\citepunct}{,\penalty\citepunctpenalty\,}
\renewcommand{\citedash}{--}

\begin{document}

\title{``\textit{Privacy is the Boring Bit}'': User Perceptions\\ and Behaviour in the Internet-of-Things}

\author{\IEEEauthorblockN{Meredydd Williams, Jason R. C. Nurse and Sadie Creese\\}
\IEEEauthorblockA{Department of Computer Science,\\University of Oxford, UK\\
\textit{\{firstname.lastname}\}@cs.ox.ac.uk}}

\IEEEpubid{978-1-5386-2487-6/17/\$31.00~\copyright~2017 IEEE}

\maketitle
\begin{abstract}
In opinion polls, the public frequently claim to value their privacy. However, individuals often seem to overlook the principle, contributing to a disparity labelled the `Privacy Paradox'. The growth of the Internet-of-Things (IoT) is frequently claimed to place privacy at risk. However, the Paradox remains underexplored in the IoT. In addressing this, we first conduct an online survey (N = 170) to compare public opinions of IoT and less-novel devices. Although we find users perceive privacy risks, many still decide to purchase smart devices. With the IoT rated less usable/familiar, we assert that it constrains protective behaviour. To explore this hypothesis, we perform contextualised interviews (N = 40) with the public. In these dialogues, owners discuss their opinions and actions with a personal device. We find the Paradox is significantly more prevalent in the IoT, frequently justified by a lack of awareness. We finish by highlighting the qualitative comments of users, and suggesting practical solutions to their issues. This is the first work, to our knowledge, to evaluate the Privacy Paradox over a broad range of technologies.
\end{abstract}


\section{Introduction}
\label{sec:one}

Opinion polls and surveys suggest that the public value their privacy \cite{Rainie2013,TRUSTe2015}. However, research indicates that individuals often act to the contrary \cite{Carrascal2013,Beresford2012}. This apparent disparity between opinions and actions has been labelled the `Privacy Paradox' \cite{Barnes2006}. While some attribute concerns to social norms \cite{Fazio2005}, others believe cognitive biases \cite{Acquisti2004a} have an influence.

The Internet-of-Things (IoT) refers to the agglomeration of `smart devices' which increasingly pervade our lives. These networks offer great benefits to productivity, being widely predicted to benefit production. However, despite the appeal of these networks, many have highlighted their threats to privacy \cite{Abomhara2014,Williams2016c}. As this field rapidly expands, what does this mean for perceptions, behaviour and the Privacy Paradox?

Thus far the Paradox has been studied in less-novel environments. Whereas the issue is explored on smartphones \cite{Park2015} and social networks \cite{Acquisti2006}, it is rarely examined in the IoT \cite{Hallam2016}. Furthermore, Paradox studies have been criticised for comparing abstract concepts with practical behaviour \cite{Trepte2014a}. Student samples are frequently solicited, with little consideration of real-life scenarios. Without practical analysis of the Paradox, the IoT might place user privacy at risk.

To explore the phenomenon across both IoT and less-novel products, we conducted two detailed studies. Firstly, to compare opinions of a range of devices, we undertook an online survey (N = 170). We sought evaluations before requesting the rationale for product ownership. IoT devices were considered significantly less private/usable, suggesting protection might be constrained. Although most users recognised the risks, many still decided to purchase IoT products. 

Intrigued by this potential disparity between opinion and action, we conducted contextualised interviews with the public (N = 40). Rather than comparing the abstract and the practical, we grounded discussions around each participant's device. 1/3 of our respondents displayed an opinion-action disparity, suggesting the presence of the Paradox. While some non-IoT owners acted in this manner, the disparity was significantly more prevalent in IoT users \cite{Williams2017}. We hypothesise this to be due to reduced awareness, with this rationale predominantly given by participants. We finally proposed solutions aligned with these justifications, such as IoT educational campaigns. \IEEEpubidadjcol

Our work is the first to analyse the Privacy Paradox across such a range of devices. We are also the first to compare privacy perceptions between the IoT and less-novel products. Rather than studying student-composed convenience samples, we dissect the privacy rationale of the general public. Our work offers novel insights into the Privacy Paradox, and provides practical solutions to reduce its prevalence.

This paper is structured as follows. Section \ref{sec:two} reviews literature on the Paradox and privacy decision-making. Section \ref{sec:three} details our methodology, before Section \ref{sec:four} reflects on the findings. Section \ref{sec:five} introduces our contextualised interviews, followed by the discussion in Section \ref{sec:six}. We conclude the paper in Section \ref{sec:seven}, highlighting limitations and further work. 

\section{Background and Related Work}
\label{sec:two}

\subsection{The Privacy Paradox}

While the principle of privacy is widespread, it is also cultural and subjective. With the concept being highly contextual \cite{Nissenbaum2009}, people might value privacy in one situation but not another. Clarke \cite{Clarke1999} defined information privacy as, ``\textit{the interest an individual has in controlling, or at least significantly influencing, the handling of data about themselves}'' \cite{Clarke1999}. While we concern this domain in our work, people might have varying views of privacy in other contexts.

Opinion polls suggest that the public care about privacy. 86\% of US respondents reported taking steps to protect themselves \cite{Rainie2013}, while 88\% in a UK study claimed to value the principle \cite{TRUSTe2015}. Despite these assertions, individuals often express behaviour to the contrary. Carrascal et al. \cite{Carrascal2013} used an auction to assess the value placed on personal data. They found users would sell their browsing history for \euro 7, suggesting a lack of concern. Beresford et al. \cite{Beresford2012} varied the prices of two online stores to explore privacy valuation. They discovered that when the intrusive store was \euro 1 cheaper, almost every user selected it. Although people might claim to be concerned about privacy, their behaviour can often appear misaligned.

This disparity between opinion and action has been labelled the `Privacy Paradox' \cite{Barnes2006}. Decision-making is also dissected through `Privacy Calculus' \cite{Dinev2006}, where disclosure benefits and risks are logically compared. However, Acquisti \cite{Acquisti2004a} prefers `bounded rationality' to explain behaviour, noting that decisions are constrained by cognitive biases. In a 2017 review, Barth and Jong \cite{Barth2017} also concluded that irrationality is present. With the phrase `Privacy Paradox' under dispute, we prefer `disparity' to describe a discrepancy between privacy opinions and actions. This is similar to the concept of the `attitude-behaviour gap' found in psychological research \cite{Fazio2005}.

\subsection{Privacy Decision-Making}

Privacy decision-making has been analysed through many studies. Acquisti and Grossklags \cite{Acquisti2005b} rejected perfect rationality, instead considering the role of cognitive biases. They conducted a 119-person survey, identifying a disparity between concerns and behaviour. As most users could not assess their risk, they concluded that a lack of awareness was influential.

Dinev and Hart \cite{Dinev2006} developed the Extended Privacy Calculus Model, analysing the balance between risks and incentives. Through surveying 369 participants, they confirmed that perceived risks led to a reluctance to disclose. Xu et al. \cite{Xu2009} investigated location-based services and the factors which influence privacy decisions. Through their survey, they found compensation increased perceived benefits while regulation reduced perceived risks. While these works are purely quantitative, we follow a mixed-methods approach. Furthermore, while they have relevance to our research, they do not concern the IoT. As this field differs in terms of usability \cite{Foster2016} and ubiquity, decisions might differ from those on familiar systems.

Although the Privacy Paradox is rarely studied in the IoT, Hallam and Zanella \cite{Hallam2016} did consider wearable devices. They constructed a self-disclosure model before validating it through an online survey. They found that behaviour was more driven by short-term incentives than long-term risk avoidance. Li et al. \cite{Li2016} studied Privacy Calculus in wearable healthcare products. Through surveying 333 users, they found that adoption increased as functionality outweighed sensitivity.

While the Paradox is not considered, other work explores IoT privacy. Wieneke et al. \cite{Wieneke2016} studied wearable devices and how privacy affects decisions. Through 22 interviews, they found individuals had little awareness of data sharing. Most also claimed risk did not impact their choices, which might suggest an opinion-action disparity. Lee et al. \cite{Lee2016b} surveyed 1,682 users on their wearable perceptions. Participants indicated their concern following privacy infractions by a hypothetical product. They found preferences correlated with reactions, even in unfamiliar situations. While these studies analyse wearable devices, we explore a variety of technologies.

Kowatsch and Maass \cite{Kowatsch2012} developed a model to predict IoT disclosure intention. They conducted surveys with 31 experts, finding usefulness the only factor to consistently encourage usage. Yang et al. \cite{Yang2017} also considered how concerns affect smart home adoption. The authors developed a theoretical model and validated it through a 216-person survey. They found that while privacy risks limit adoption, trust can counteract the effect. Whereas these studies explore a few scenarios, we examine a range of IoT and less-novel products. This enables analysis of how technology influences behaviour. With the IoT proliferating, it is crucial we ascertain its influence on privacy.

\section{Online Survey Methodology}
\label{sec:three}

\subsection{Research Hypotheses}

Before we describe our methodology, we must outline our research hypotheses. These can be found below in Table \ref{tbl:hypotheses1}. They were based on our research goal: to explore the Paradox across IoT and less-novel environments. To ascertain high-level opinions, we designed a public online survey. Rather than solely analysing privacy, we explored other factors which could have an influence. For example, less usable or (less) familiar devices might constrain protective behaviour \cite{Whitten1999}. Therefore, as outlined in Section \ref{sec:evalfact}, we asked respondents to evaluate four factors: \textit{privacy}, \textit{familiarity}, \textit{usability} and \textit{utility}. This enabled us to compare opinions of IoT and non-IoT products, with device selection described in Section \ref{sec:devsec}.

\begin{table}[h!]
\caption{Online Survey Research Hypotheses}
\setlength{\tabcolsep}{.35em}
\begin{tabular}{cl} 
\toprule
\# & Research Hypothesis \\ 
\midrule
H1 & \pbox{10cm}{Mean privacy ratings for IoT devices will be significantly less than \\ mean privacy ratings for non-IoT devices.}
 \\[2.5ex]
H2 & \pbox{10cm}{Mean familiarity ratings for IoT devices will be significantly less \\ than mean familiarity ratings for non-IoT products.}
 \\[2.5ex]
H3 & \pbox{10cm}{Mean usability ratings for IoT devices will be significantly less than \\ mean usability ratings for non-IoT products.}
 \\[2.5ex]  
H4 & \pbox{10cm}{Participants will be significantly more likely to own an IoT device \\ while giving it a low privacy rating (less than 2/5) than to own \\a non-IoT device and give it a low privacy rating.}
 \\
\bottomrule
\end{tabular}
\label{tbl:hypotheses1}
\centering
\end{table}

Since studies suggest smart devices could impact privacy \cite{Abomhara2014,Williams2016c}, we believed non-experts would share this opinion. We therefore asserted that IoT products would be rated less privacy-respecting than non-IoT technologies (\textbf{H1}). With smart devices being heterogeneous \cite{Bandyopadhyay2011} and novel, we posited these technologies would also be less familiar (\textbf{H2}). This has particular risk for privacy behaviour, as users might be less able to use protection \cite{Whitten1999}. As IoT interfaces are often criticised \cite{Foster2016}, we asserted they would be rated less usable (\textbf{H3}).

Following the factor ratings, we queried participants on whether they owned the device and why. This qualitative justification sought to identify factors influencing ownership decisions. While we believed the IoT would be considered less private, we doubted this would reduce its popularity. Therefore, we posited that this disparity between opinion and purchasing action would be more prevalent in the IoT (\textbf{H4}).

\subsection{Survey Design}

We chose to begin with an online survey, enabling the analysis of public opinion. Being directed by our high-level findings, we then explored in depth through qualitative interviews. The questionnaire was advertised via Twitter and national/international message boards. Such boards included DailyInfo, GumTree and The Student Room. These fora were selected as we wished to canvas non-expert opinions. No screening criteria were applied, other than the participants being adults. The questionnaire was iteratively refined, with face validation received from privacy and psychometric experts. We sought to disguise an IoT/non-IoT comparison, framing the theme as general technology. We then performed a small pilot test, before the survey was undertaken from Sept to Nov 2016. The form was composed of demographics and factor ratings, with these components discussed in the following subsection.

\subsection{Demographics and Factor Ratings}
\label{sec:evalfact}

We solicited gender, age and highest education level. As research \cite{Sheehan1999} suggests women possess larger privacy concerns than men, we explored whether privacy ratings varied similarly. It is also reported that older people care more about privacy \cite{Han2002}, and this could be reflected in conservative evaluations. Previous work found that education correlates with privacy concern \cite{ONeil2001}, and this could influence our ratings.

Ratings were made from 0 (low) to 5 (high) on an ordinal scale. This scale was selected for simplicity to aid our non-expert audience. As previously mentioned, these factors were \textit{privacy}, \textit{familiarity}, \textit{usability} and \textit{utility}. We chose these non-privacy attributes both to disguise survey purpose and for their aforementioned interest to the study. We chose against including factor definitions, as we wished to explore the unbiased opinions of our non-expert participants. We did substitute `utility' for `usefulness' on the form, as we believed this synonym to be more comprehensible.

\subsection{Device Selection}
\label{sec:devsec}

Through our above factors, we compare smart devices with less-novel alternatives. However, with the IoT being nebulous, we constrained our scope. We chose to select six technologies: three IoT and three non-IoT. These labels are not a strict dichotomy; there is a spectrum ranging from novel mobile products to familiar desktop computers. However, to enable a comparison between groups, we selected archetypal products.

Since we sought public opinion, we constrained our focus to consumer devices. We then specified three criteria to aid selection: novelty, ubiquity and autonomy. These were chosen as IoT products are typically modern, ubiquitous and autonomous. Whereas PCs are well-established, the IoT has flowered in the past decade (novelty). Although laptops reside in many houses, they do not pervade like `smart homes' (ubiquity). Finally, older products are typically user-dependent, while the IoT often interacts with its surroundings (autonomy).

By plotting products against novelty, ubiquity and autonomy, we identified which devices fell into which group. Desktops and laptops appeared non-IoT: both are over 20 years old; both require input; and neither would be considered a Ubicomp device. While tablets do have greater portability, they possess similarities to a keyboardless laptop. Since technology research firms \cite{Rivera2013,Duffy2014} also judge these products as distinct from the IoT, we are confident in our categorisation. Furthermore, smart products often require human-free interaction \cite{Rouse2016}, which is rarely supported by desktops, laptops or tablets. 

Wearables (e.g., Fitbit) have achieved recent success, are highly mobile and use autonomous sensors. Smart appliances, such as connected fridges, are also novel and communicate through online interaction. Home automation systems (e.g., Google Nest), while static, are highly pervasive and react to their environments. We therefore compared (\textit{desktops, laptops, tablets}) with (\textit{wearables, smart appliances, home automation}). Although definitions were not provided (as a means of disguising the IoT/non-IoT comparison) we included images of relevant devices. These products originated from a range of manufacturers to reduce bias from brand predilections. While some products sit between categories, such as the Microsoft Surface, such examples are rare. Furthermore, although diversity exists within groups, the distinction across categories is generally greater. While a Fitbit differs from an Apple Watch, they largely support similar functionality.

\subsection{Response Bias Mitigation}

Since self-reporting surveys face a number of risks, we sought to mitigate response biases. Privacy concerns can become inflated if the topic is salient \cite{Rajivan2016}. Therefore, we disguised the subject through a generic survey with a range of factors. As acquiescence bias can lead participants to agree with researchers, we avoided yes/no questions. While later ratings might be made relative to earlier scores, we shuffled categories to mitigate the effect. To both allay concerns and reduce non-response bias, we treated data anonymously and received ethical approval. This was important, as otherwise those most concerned about privacy might avoid the study.

\section{Survey Results and Discussion}
\label{sec:four}

\subsection{Participants and Techniques}

We collected 170 responses with 57\% male and 43\% female. 50\% came from the 26-35 age group, reflected in our estimated mean of 32. 36\% of participants possessed a Master's degree, implying a well-educated respondent group.

For correlation, we analysed the Spearman's Rank-Order Correlation Coefficient ($r_s$). We used this technique as our variables were ordinal and varied monotonically. To perform significance testing on two independent samples, we used the Mann-Whitney U Test. We selected this as our dependent variables were ordinal and our independent variables were nominal. When comparing two related samples, we chose the Wilcoxon Signed-Rank Test. This was used as our dependent variables were ordinal and our independent variables had related groups. In all cases, we required \textit{p}-values \textless \xspace 0.05 for significance. We discuss three opinion variables: the mean \textit{privacy} rating, the mean \textit{familiarity} rating and the mean \textit{usability} rating. We use $\bar{x}$ to denote means, $r_s$ for correlation coefficients and include \textit{p} when differences are significant.

\subsection{Demographic Analysis}

While women rated technologies less privacy-respecting than men, this difference was not significant. This might have been due to our sample size, or because opinions are unformed for unfamiliar products. We found age was significantly negatively correlated with privacy ratings ($r_s$ = -0.232, \textit{p} = 0.002), implying older people express greater concern. This might be a generational issue, as older individuals did not grow up with new devices. While we found the highly-educated did rate products less privately, the correlation was not significant. Again, this could be due to the unfamiliarity of IoT products.

\subsection{Factor Comparisons}

To understand how opinions differ, we compared factor ratings across our surveyed devices. We first calculated the mean score of each factor for each product. We then performed significance testing to investigate our hypotheses.

Laptops ($\bar{x}$ = 3.27) were rated most private, with wearables ($\bar{x}$ = 2.31) regarded as most privacy-concerning. Generally, we found IoT devices were rated significantly less privacy-respecting (\textit{Z} = -5.151, \textit{p} \textless \xspace 0.001), confirming our hypothesis (\textbf{H1: Accept}). This might be due to fear of the unknown, or because IoT products collect a range of data. If the public recognise the risk, it implies security awareness is increasing.

Laptops ($\bar{x}$ = 4.72) were also found most familiar, with home automation receiving the lowest score ($\bar{x}$ = 1.45). Individuals were significantly less-accustomed to IoT devices (\textit{Z} = -11.103, \textit{p} \textless \xspace 0.001), which we imagine to be due to their current novelty (\textbf{H2: Accept}). Low familiarity could lead to privacy risks, with users less aware of protective techniques \cite{Whitten1999}. If these technologies are considered alien, this implies that the IoT market is still nascent. We would expect such devices to become better-known as their sector expands.

Laptops were considered most usable ($\bar{x}$ = 4.55), with wearables faring the worst ($\bar{x}$ = 2.52). IoT products were considered significantly less-usable than their counterparts (\textit{Z} = -10.332, \textit{p} \textless \xspace 0.001), confirming our hypothesis (\textbf{H3: Accept}). This could be because the gadgets are less-understood, or because they often possess small screens. With the IoT rated less usable and (less) familiar, protection might be impeded \cite{Whitten1999}. The mean factor ratings are presented below in Figure \ref{fig:chart}.

\begin{figure}[h!]
    \includegraphics[width=0.46\textwidth]{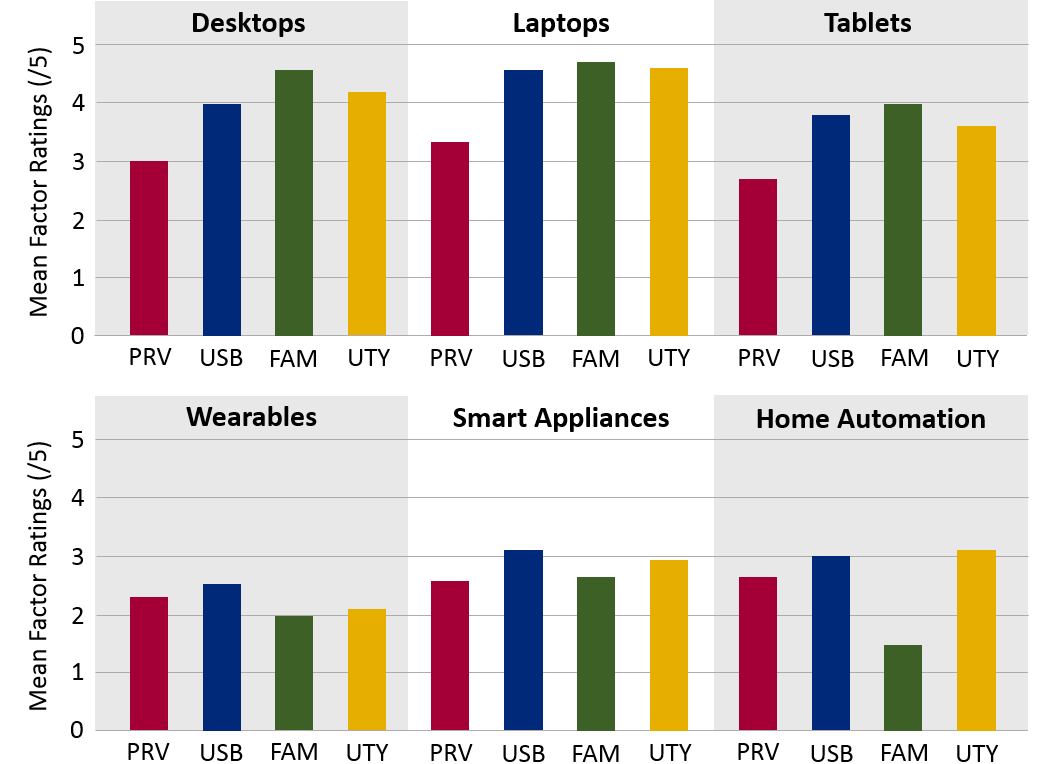}
    \centering
    \caption{Device mean factor ratings: Privacy (PRV), usability (USB), familiarity (FAM) and utility (UTY).}
    \label{fig:chart}
\end{figure}

\subsection{Ownership Decisions}

With IoT products considered less-private (H1), we investigated whether this affected purchasing decisions. For each device, we compared privacy ratings to ownership frequency and user justifications. If a person recognises the risks but still purchases the product, then a disparity might be present.

Laptops were most popular, with 93\% claiming ownership. Mobility was the most liked feature, with no privacy concerns expressed. Tablets were owned by 66\%, with ownership mainly justified on mobility. Only one person criticised privacy, with them denouncing a lack of settings customisation. Desktops were owned by 52\%, with 62\% of those praising the product's functionality. Again, not a single person criticised privacy. This implies that the topic is rarely influential when users purchase computers.

Smart appliances were owned by 42\%, with functionality the most popular feature. 9\% rejected because of privacy, with these participants worried about monitoring. Despite this, more than 3 times as many were deterred by price (28\%). Only 21\% owned wearables, with functionality again the main attraction (63\%). Although they received the lowest privacy ratings, only 3\% of rejections cited this reason. Again, far more were deterred by the cost (20\%). Of those 12\% with home automation, only 8\% of them criticised privacy. While they disliked remote infiltration, again, far more blamed the price (23\%). Since prices decrease as products mature, this suggests the IoT will grow in popularity. Therefore even if privacy becomes salient, cheap gadgets might remain attractive.

\subsection{The Opinion-Action Disparity}
\label{sec:opacsurv}

We now move forward to compare privacy ratings with purchasing decisions. If individuals buy a device despite recognising the risks then the disparity might be present. We take ratings of 1/5 or below to indicate criticism, as 2/5 (or 3/5) could be deemed a cautious evaluation. In this manner, we seek to place a minimum bound on disparity prevalence.

In the case of non-IoT technologies, the disparity was far from common. Of the 89 who bought a laptop, only 7 gave a low privacy rating (7.87\%). For tablets the figure was 10.71\%, with desktop disparities even less common (8.23\%). On average, 8.91\% purchased a non-IoT product despite perceiving a risk. These individuals may have felt constrained by the PC market, with operating systems developed by a small number of vendors. Therefore, even if users object to a brand's privacy practices, they have few alternatives to choose from.

The disparity was more prevalent for IoT devices. Of those who purchased home automation systems, almost 10\% rated privacy poorly. 9/71 smart appliance owners criticised privacy (12.68\%), with wearables performing the worst (17.14\%). Across all IoT owners, this resulted in an average of 14.96\%. These purchases might be made because consumers value functionality over data privacy. Alternatively, owners might have inconsistent preferences and detach their opinions from their actions. In either case, this implies that a subset of individuals are willing to sacrifice their privacy. This is concerning as while employees might require a desktop/laptop for work, IoT purchases are largely voluntary. In this manner, privacy is sacrificed for entertainment rather than necessity. Although IoT users were 68\% more likely to present this disparity, there was no significant difference between our groups (\textbf{H4: Reject}). With a \textit{p}-value of 0.056, our sample may have been too small to confirm the difference.

\subsection{Findings and Implications}

We found the IoT is regarded as less privacy-respecting (H1), less familiar (H2) and less usable (H3). Since confusing and unfamiliar interfaces are harder to use, data protection might be impeded \cite{Whitten1999}. Ownership justifications imply privacy is rarely considered, and this could contribute to unwise purchases. If the topic is not salient, users might place themselves at risk. While our findings hinted at constrained action, this could not be confirmed without additional data. As decision-making was opaque in these quantitative results, we required detailed discussions. We therefore undertook qualitative analyses to dissect the decision-making rationale. In this manner, we can compare opinion and action to explore the disparity.

\section{Contextualised Interview Methodology}
\label{sec:five}

\subsection{Research Hypotheses}

Again, we begin this section by introducing our hypotheses. These can be found below in Table \ref{tbl:hypotheses2}. As described in the following sections, we conducted contextualised interviews with a non-expert public. In these discussions, we solicited both participants' opinions and their reported behaviour. These responses were then codified and quantified, resulting in the metrics described in Section \ref{sec:interviewdataanalysis}. This enabled direct comparison between individuals' privacy opinions and their actions. In this manner, the prevalence of the disparity could be evaluated across both IoT and less-novel environments.

\begin{table}[h!]
\caption{Contextualised Interview Research Hypotheses}
\setlength{\tabcolsep}{.35em}
\begin{tabular}{cl} 
\toprule
\# & Research Hypothesis \\ 
\midrule
H5 & \pbox{10cm}{Mean quantified privacy opinions will be significantly less for IoT \\ devices than for non-IoT devices.}
 \\[2.5ex]
H6 & \pbox{10cm}{Mean quantified privacy actions will be significantly less for IoT \\ devices than for non-IoT devices.}
 \\[2.5ex] 
H7 & \pbox{10cm}{Disparities (two-point gaps between quantified privacy opinion and \\ quantified privacy action) will be significantly more likely for IoT \\users than non-IoT users.}
 \\
\bottomrule
\end{tabular}
\centering
\label{tbl:hypotheses2}
\end{table}

As more IoT users bought `risky' devices, we posited such owners would care less about their data. Furthermore, since privacy rarely featured in ownership decisions, this suggests the topic is infrequently considered. We asserted that this would result in lower quantified opinion scores (\textbf{H5}), with individuals showing less concern. This comprised the opinion component of the opinion-action disparity. As our survey suggested the IoT is less familiar (H2) and (less) usable (H3), we posited users would also be less able to protect themselves. Smart devices are heterogeneous and novel, potentially challenging mental models. This might result in lower quantified action scores (\textbf{H6}), with users doing less to protect themselves. This comprised the action component of the disparity.

As IoT owners displayed the disparity frequently in the survey, we asserted that it would be prevalent in the interviews. While their privacy expectations might be lower, we contend that their behaviour is disproportionately constrained. If true, this would lead to an increased discrepancy between perceptions and behaviour. We therefore posit that the opinion-action disparity (the Privacy Paradox) will be more likely in the IoT (\textbf{H7}), with metrics outlined in Section \ref{sec:interviewdataanalysis}.

\subsection{Interview Design}

With our survey suggesting a potential disparity, rich data was required for further investigation. Therefore we designed interviews to discuss privacy rationale. As we wished to explore the wider applicability of the disparity, we approached a distinct sample of the general public. If a subset of these also display the disparity, then the phenomenon might be common.

To compare IoT and non-IoT owners, we recruited two distinct groups. These were divided based on the survey categories, enabling analysis of whether the same dichotomy exists. Both groups faced the same questions, with only the device name customised in our between-subjects format. Participants were screened for adults who owned a device in one of our six categories. Recruitment was undertaken via Twitter and a local messaging board, ensuring our respondents did not comprise a student sample. Interviews were conducted one-on-one in a seminar room, with informed consent received at the start. Monetary compensation was offered to incentivise participation, and the study was approved by our IRB board.

If participants believe their privacy perceptions are being evaluated, they might adjust their responses \cite{Rajivan2016}. Therefore, our interview was framed as concerning general opinions. More-overt questions were also placed near the end to ensure earlier responses were not primed. To minimise any deception, the true purpose was revealed at the end of each interview.

We sought to overcome the criticisms of previous Privacy Paradox studies \cite{Trepte2014a}. Firstly, rather than comparing abstract concepts against practical actions, we grounded our interviews around owned devices. Participants were then able to draw on their personal experiences to answer in a more-informed manner. With privacy being highly contextual, this enabled opinions and actions to be fairly compared. Secondly, instead of considering `privacy', we solicited qualitative reactions to specific incidents. Rather than discussing the nebulous principle, as has been criticised by previous work \cite{Dienlin2015}, we constrained our focus to informational privacy. Thirdly, our interviews were conducted with the public, as opposed to student-composed samples. This should lead our findings to be more-representative of non-expert users.

Fourthly, we discussed protective actions (described below) that were both practical and feasible. While few non-experts use Tor, passwords and settings can help ordinary users. Finally, we considered the rationale behind decisions, rather than just the decisions themselves. If a password is neglected because the data is thought trivial, then the user is not necessarily careless. If despite these controls, they express concern but take no action, we argue a disparity is present.

\subsection{Interview Questions}

We first received face validation from a privacy and psychometric expert. We then conducted a pilot study, granting an opportunity to test our questions. Following interviews with 10 individuals, we found our sequence primed privacy. After moving our action queries to later in the session, the topic appeared better disguised. While our interview questions were broad, they were chosen to solicit open-ended comments from non-expert users. A more prescriptive approach might have channelled responses, but also constrained the diversity of replies. Privacy Paradox studies have been criticised for comparing abstract opinions with specific circumstances \cite{Trepte2014a}. For example, while a person might value privacy, this may bear no relation to their Facebook usage. To ensure that opinions and actions are comparable, we contextualised our questions around a participant's device. For example, if they owned a Fitbit, all queries concerned their use of that Fitbit.

Questions were of four types: \textit{General (G)}, \textit{Opinion (O)}, \textit{Action (A)} and \textit{Disparity (D)}. These queries can be found below in Table \ref{tbl:questions}. \textit{General} questions had two roles: to solicit broad opinions and to disguise the topic of privacy. Although our \textit{General} questions led to intriguing findings, in the interest of brevity, we scope to our other results. \textit{Opinion} queries were used to investigate privacy perceptions. Incidents were selected from the archetypal privacy violations found in Solove's taxonomy \cite{Solove2008}. Disclosure and surveillance are both comprehensible ways in which privacy can be violated. Data selling encapsulates the secondary use violation, while unauthorised deletion represents an intrusion into solitude.

\begin{table}[h!]
\caption{Contextualised Interview Questions \protect\linebreak (General Questions excluded for brevity)}
\setlength{\tabcolsep}{.35em}
\begin{tabular}{cl} 
\toprule
\# & Interview Question \\ 
\midrule
O1 & \pbox{10cm}{How would you feel if someone deleted your X's data without your \\ permission? Why?}
 \\[2ex] 
O2 & \pbox{10cm}{How would you feel if someone shared your X's data without your \\ permission? Why?}
 \\[2ex]
O3 & \pbox{10cm}{How would you feel if someone monitored everything you do on \\ your X? Why?}
 \\[2ex]
O4 & \pbox{10cm}{How would you feel if someone sold your X's data without your \\ permission? Why?}
 \\[2ex] 
A1 & \pbox{10cm}{Does X allow you to set a password? Have you set a password? \\Why (not)?}
 \\[2ex] 
A2 & \pbox{10cm}{How much time have you spent reading X's privacy policies? Why?}
 \\[1ex]
A3 & \pbox{10cm}{How much time have you spent configuring X's privacy settings? \\Why?}
 \\[2ex]
D1 & \pbox{10cm}{Why do you think some people use devices which place their \\privacy at risk?}
 \\[2ex] 
D2 & \pbox{10cm}{Why do you think some people use their devices in an unprivate \\way?}
 \\[2ex]
D3 & \pbox{10cm}{Why do you think some people claim to value privacy but still use \\ devices which place their privacy at risk?}
 \\
\bottomrule
\end{tabular}
\centering
\label{tbl:questions}
\end{table}

\textit{Action} questions queried how participants actually use their devices. Protective measures were selected based on three criteria: simplicity, utility and applicability. Techniques must be easy to apply, as we should not expect non-experts to install complex software. Measures must also be beneficial by granting an opportunity for greater knowledge or control. Finally, techniques must apply to both IoT and non-IoT devices to enable a fair comparison. Passwords, privacy policies and privacy settings are all of use, widespread and well-known. Therefore, we avoid comparing opinions against impractical actions. While opaque policies frequently lack usability, they still offer an opportunity to discover device practices. 

In addition to assessing the disparity's existence, we explored privacy rationale. To avoid priming the topic, the \textit{Disparity} questions were placed at the end of the interview. We believed disparity-prone individuals might respond defensively if directly queried on the topic. Therefore, we phrased questions in terms of why other people might act in this manner. While answers were likely to still correspond with their rationale, we avoided antagonising our respondents.

\section{Interview Results and Discussion}
\label{sec:six}

\subsection{Participants}

We conducted 40 contextualised interviews between January and February 2017. 60\% were male and 40\% were female, closely corresponding with the 57\%/43\% split in our survey. Respondent ages were also similar, with 45\% in the 26-35 group and an estimated mean of 31.6. Educational levels were again relatively high as 53\% possessed a Master's degree.

\subsection{Data Analysis}
\label{sec:interviewdataanalysis}

We manually transcribed our recordings, resulting in a transcript for each discussion. We then conducted thematic analysis, labelling responses under a range of codes. Rather than simply noting the answer, these codes also encapsulated the justification for the decision. After all transcripts were reviewed, categories were developed to ensure consistency. For example, privacy policy codes `\textit{Did Not Read, Jargon}' and `\textit{Did Not Read, Legalese}' were categorised under `\textit{Did Not Read, Complex}'. Where justifications were clearly distinct, they were preserved to ensure a diversity of views.

In terms of opinions, the intensity of reaction was grouped under `\textit{Indifference}', `\textit{Slight Dislike}', `\textit{Dislike}' or `\textit{Strong Dislike}'. While a `\textit{Like}' category was envisaged, none of the participants expressed this reaction. These groups were used to distinguish between those who felt inconvenienced and those who showed strong opposition. Actions were split between `\textit{Did}' and `\textit{Did Not}' unless a ordinal scale appeared necessary. For example, as a sizeable proportion of participants skimmed their policies, responses were divided ordinally between `\textit{Did}', `\textit{Briefly}' and `\textit{Did Not}'. In seeking to minimise subjectivity, categorisation was refined to ensure group consistency.

To assess the disparity at an individual level, we quantified opinions and actions. Whereas a comparison could be made qualitatively, we believed this approach would be too subjective. Opinions were scaled from 1/5 (low) to 5/5 (high) based on concern intensity and justification. For example, a person with `\textit{Strong Dislike}' towards deletion, surveillance and selling would receive 5/5. If concerns were contingent on a particular factor, such as high sensitivity, the score was reduced. 

We used a similar scale for privacy actions, assessing whether participants set passwords, read policies and configured settings. Their rationale was also considered, as a person might reject a password for trivial data. In these cases their action score was increased. These adjustments sought to place a minimum bound on disparity prevalence.

To identify disparities, we judged whether the opinion and action scores were commensurate. As both question sets were contextualised around the same device, we ensured a correspondence between scores. Furthermore, the actions could be used to directly address the hypothetical violations. For example, passwords can reduce deletion risk, while policies outline how devices are monitored. If users claim concern but take little protective action, a disparity might exist.

Considering the 5-point scale, we defined a disparity as when the action score was at least 2 points less than the opinion score. We did not believe 1-point differences signified a dissonance, but thought a 3-point definition was too extreme. If a respondent strongly objects to threats (5/5) but merely glances at policies and settings (3/5), then their behaviour might be deemed unwise. Similarly, if a person exhibits reasonable concern (3/5) but takes no action (1/5), then they might also be at risk. As we controlled for contingent concerns, we placed a minimum bound on disparity prevalence.

We continued to use the Mann-Whitney U Test to compare ordinal variables between our participant groups. When responses were binary (nominal), such as `\textit{Set Password}' and `\textit{Did Not}', the Chi-Square Test was used instead. When analysing the correlation between ordinal data, we continued to study the Spearman's Rank-Order Correlation Coefficient ($r_s$). In all cases we required \textit{p}-values \textless \xspace 0.05 for significance. As we compare distinct variables once each, we do not expect to be affected by the Multiple Comparisons Problem.

\subsection{Participant Opinions}

Opinion questions concerned data deletion, unauthorised sharing, surveillance and data selling. Most participants objected to deletion, with 73\% expressing a dislike for the scenario. This implies that individuals generally feel some sense of ownership over their data. However, we discovered IoT product owners cared significantly less about the issue (\textit{U} = 121, \textit{p} = 0.033). Smart device data was often perceived as low in value (expressed by participants including \#18, \#31 and \#34), as shown below. This is concerning, as while some data is trivial, home occupancy metrics can be revealing. Furthermore, GPS data from wearables might reveal where a person lives or works.

\begin{quoting}
``\textit{I wouldn't be too fussed, there isn't a whole lot on there that I'm particularly dear to. It's just settings and stuff like that, nothing to worry about}'' (\#34, IoT)
\end{quoting}

In terms of unauthorised sharing, 78\% either disliked or strongly disliked the practice. This implies that despite the popularity of sharing content, people want agency over this process. IoT owners cared significantly less about unauthorised sharing (\textit{U} = 81, \textit{p} = 0.001), suggesting a dichotomy in privacy opinions. Smart device users often cited a lack of data sensitivity (\#4, \#21, \#35), whereas non-IoT owners were troubled by an absence of control (\#12, \#17, \#37). While IoT metrics might not appear sensitive, users may not have knowledge of advanced inference techniques \cite{Creese2012}.

\begin{quoting}
``\textit{Just because it's only activity, it's only what I get up to, I don't see it as a secret}'' (\#35, IoT)
\end{quoting}

Both groups strongly rejected surveillance, with 85\% of IoT device users objecting to monitoring. This implies that consumers still criticise the notion of supervision. This is in conflict with modern wearables, as many of these track GPS. Whereas many non-IoT respondents rejected surveillance on principle (\#2, \#9, \#14), IoT users expressed some concern over tracking (\#15, \#23, \#35). With many smart devices offering location services, digital stalking can be a real possibility.

\begin{quoting}
``\textit{I'd feel like, like someone would maybe be stalking me which would be a bit unnerving}'' (\#35, IoT)
\end{quoting}

Data selling was also met with widespread condemnation. 83\% at least disliked the practice, with 30\% expressing strong objections. Despite the prevalence of data markets, this implies consumers still reject this custom. With information frequently sold by technology firms, users might be unaware how common this  practice is. Whereas non-IoT participants were concerned by a lack of consent (\#8, \#19, \#32), smart device users wanted money from the transaction (\#5, \#18, \#36). This suggests IoT owners have a greater understanding of how data is monetised. 

\begin{quoting}
``\textit{I would also be angry because I should get part of the share of the money}'' (\#36, IoT)
\end{quoting}

We found that 60\% expressed strong privacy opinions, being scaled to either 4/5 or 5/5. This implies that the public still claim to value this threatened principle. However, IoT users were found to have significantly lower privacy concerns (\textit{U} = 127.5, \textit{p} = 0.049) (\textbf{H5: Accept}). This confirms our hypothesis that IoT owners appear to care less about their data. From our qualitative justifications, this often appears due to the data being considered less important. Although data can appear trivial, users might not understand the inferences that can be made \cite{Creese2012}. Therefore, non-expert owners might unwittingly place their privacy at risk.

\subsection{Participant Actions}

Since our survey suggests the IoT is less familiar (H2) and (less) usable (H3), user protection might be constrained. Privacy behaviour was gauged on whether users set passwords, read their devices' privacy policies and configured their devices' privacy settings. If an individual reads their policies and adjusts their settings, they arguably behave more privately than someone who ignores these opportunities.

Password protection was far from perfect, with only 58\% securing their products. We found passwords were used significantly less often on smart devices ($X^2$(1) = 14.11, \textit{p} \textless \xspace 0.001). With these products usually connected to the Internet, users might place their data at risk. Inconvenience played a large role, with PINs often reducing usability (\#13, \#20, \#21). This justification indicates that while users might know about passwords, they opt for convenience. This presents a direct trade-off between utility and privacy/security. Modern wearables also face an increasing theft risk, and unsecured interfaces will only encourage this threat.

\begin{quoting}
``\textit{I just want to swipe it, yeah. It just takes too much time to get in there}'' (\#21, IoT)
\end{quoting}

In general, only 13\% studied privacy policies in detail, with 65\% avoiding the text. This implies a large number of consumers are held to terms of which they have no knowledge. Users might criticise practices as unconsented, but actually agree to them through opaque policies. Again, we found the IoT group was significantly less interested in the documents (\textit{U} = 117.5, \textit{p} = 0.041). Smart device owners often found functionality more exciting, deeming policies a low priority (\#23, \#28, \#40). Such an attitude contributes to users being unaware of data collection. If consumers are preoccupied with novel features, they might treat privacy as an afterthought.

\begin{quoting}
``\textit{I think I was more in a hurry to get it out of the box and set up and start using it}'' (\#40, IoT)
\end{quoting}

Settings adjustment was varied, with 35\% fully configuring and 33\% taking no action. This implies that while some are eager to adjust their devices, many rely on defaults. Displaying the contrast between groups, IoT users also configured their settings significantly less often (\textit{U} = 127.5, \textit{p} = 0.049). This was frequently justified through a lack of awareness (\#21, \#23, \#25), or because product functionality was considered more exciting (\#27, \#29, \#36). Once again, this displays a trade-off between privacy and utility. Since default settings are often permissive, IoT users might be leaking data. If individuals perceive privacy as boring, they may avoid protection and place themselves at risk. 

\begin{quoting}
``\textit{I just want to explore the functions and interesting bits not the privacy bit, privacy is the boring bit}'' (\#36, IoT)
\end{quoting}

Of the 40 participants, 68\% had their actions scaled to 3/5 or above. This was greater than the 60\% for opinions, suggesting that some users are more private than they claim. However, the IoT mean was again significantly lower than that of the non-IoT group (\textit{U} = 79.5, \textit{p} = 0.001) (\textbf{H6: Accept}). This confirms our hypothesis that IoT owners do less to protect their data. Our justifications suggest this is due to both a preoccupation with functionality and a lack of awareness. As highlighted in  B\"uchi et al.'s 2017 work \cite{Buchi2017}, if users do not understand protection, then they cannot guard their data.

\subsection{The Opinion-Action Disparity}

While the IoT group does less on average, we must consider individual cases to identify disparities. 13/40 participants displayed a 2-point difference between opinion and action (33\%). With these individuals recruited from a non-expert general public, this implies that the disparity might be prevalent. Furthermore, with our sample disproportionately-educated, this may be a minimum bound. While 23\% of these owned less-novel devices, 77\% possessed IoT products. Accordingly, we found IoT owners are significantly more likely to display the disparity ($X^2$(1) = 5.584, \textit{p} = 0.041) (\textbf{H7: Accept}). This confirms our hypothesis and suggests that IoT products might exacerbate the Privacy Paradox. If so, privacy might be placed at risk as smart devices proliferate. The distribution of opinions and actions are displayed below in Figure \ref{fig:map}. As the figure suggests, IoT users are more likely to have strong concerns but take little action. With the most protected participants using non-IoT products, smart devices may be a constraint.

\begin{figure}[h!]
    \includegraphics[width=0.46\textwidth]{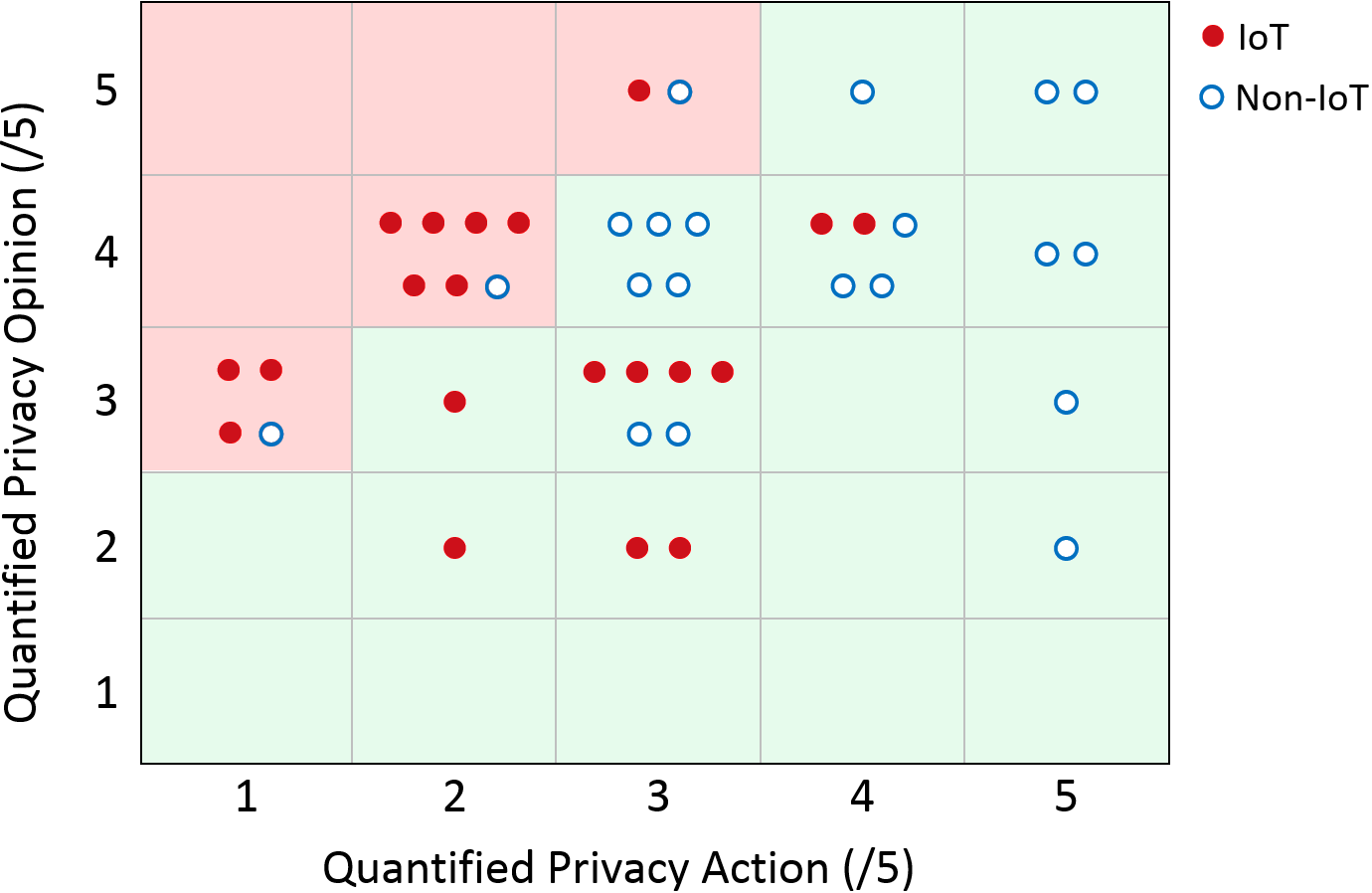}
    \centering
    \caption{Participant privacy opinion-action distribution: The shaded red area highlights where there is a disparity between privacy opinion and action.}
    \label{fig:map}
\end{figure}

If these technologies are more likely to support the disparity, why is this the case? With both concern and protective behaviour reduced in the IoT, one would expect a similar disparity prevalence. However, although IoT users often regarded their data as trivial, they still objected to privacy violations. In seeking to mitigate the issue, we explored rationale in greater detail. Through our final three \textit{Disparity} questions, we triangulated why individuals might act in this manner. Although these queries referred to other people, 77\% of disparity-prone respondents made reference to themselves.

If individuals are not aware their privacy is at risk, they cannot protect themselves \cite{Acquisti2005b}. Even if they have some knowledge of the threat, they cannot guard their data if they cannot use the system. Disparity-prone participants cited lack of awareness six times (\#19, \#21, \#27, \#32, \#33, \#35), as did 28/40 respondents. Representative quotes are presented below:

\begin{quoting}
``\textit{If that was me, I wouldn't realise until somebody said `you do realise that this is open to everybody', I'd be like `oh no' and I would change it}'' (\#21, IoT)
\end{quoting}

\begin{quoting}
``\textit{And certainly just undertaking this interview highlighted to me in ways which I may be risky}'' (\#27, Non-IoT)
\end{quoting}

It is often considered a social norm to value privacy, whether or not one's actions match their claimed concern \cite{Fazio2005}. Even if one does not care about their data, there is social pressure to desire privacy. When participants were asked why the disparity exists, social norms were mentioned most frequently (12/40). This reason was salient with disparity-prone users, mentioned by the 33\% who referred to themselves (\#1, \#23, \#25).

\begin{quoting}
``\textit{There's certainly a cultural norm of saying privacy is important, which maybe doesn't always translate into reality or action}'' (\#23, IoT)
\end{quoting}

\begin{quoting}
``\textit{Well its socially unacceptable to say `oh I don't care about privacy at all', and therefore you want to say that you do care about privacy, but in fact you're not doing very much}'' (\#25, IoT)
\end{quoting}

Individuals might understand the risks of an action, but do it anyway due to short-term necessity \cite{Hallam2016}. For example, although public Wi-Fi can be insecure, a person might still use it to send an urgent email. Security fatigue \cite{Furnell2009} describes the cognitive load users face in following security, and a similar concept might exist for privacy. While privacy can still be aspired to as a principle, it is often sacrificed due to practical necessity. This justification was offered frequently by our respondents (\#11, \#18, \#40).

\begin{quoting}
``\textit{If you need a service and you're in a rush and you need to get something done really quickly, you don't really give a s**t about the privacy bit}'' (\#18, IoT)
\end{quoting}

\begin{quoting}
``\textit{If you're travelling, sometimes you might have to use a laptop like in a caf\'{e} or in a hotel or something like that which I always try not to do, but I think that's just what makes people do it, the need to do it}'' (\#40, IoT)
\end{quoting}

\subsection{Participant-Informed Solutions}

We have identified several justifications for disparity-prone behaviour. If actions are not commensurate with opinions, users might place themselves at risk. With a third of our sample acting in this manner, further work is required. Therefore, we suggest approaches directly informed by disparity-prone individuals. We are cognisant that technology firms might resist change, as they profit from data monetisation \cite{Robinson2015}. With this in mind, we give balanced feasible suggestions.

Many respondents (\#35, \#36, \#37, \#38, \#39) recommended awareness campaigns as a means of increasing understanding. While initiatives have frequently concerned security \cite{Bada2015}, few have specifically targeted smart devices. Sessions could be held for school pupils, as they will mature in a connected world. For an effective initiative, topics including default settings and data markets must be addressed. Practical advice would be essential, such as how to disable GPS tracking. If users can understand why their data is collected, they can make decisions in an informed manner. To ascertain whether such initiatives are successful, attendees could be evaluated through a longitudinal process. Whereas education far from guarantees action \cite{Bada2015}, it would give people the tools to guard their data. 

With privacy policies often long and complex, respondents appealed for simplification (\#8, \#30, \#40). If individuals could understand how their data was used, perhaps they would make prudent decisions. While attempts have been made to simplify policies, vendors are keen to resist these efforts. As an accommodation, graphical icons could be introduced to highlight functionality. A Wi-Fi symbol could denote wireless, while a padlock could represent password protection. IoT vendors could subscribe to this scheme and compete based on their functionality. Whereas consumers would still favour exciting features, privacy would not be hidden. To assess whether standards improve, icon distribution could be observed over time. Although this approach might hamper IoT innovation, it would reduce the risk of insecure infrastructure.

Several participants believed companies should do more to protect their customers (\#36, \#21). Some complained that privacy is hidden (\#37), while others argued for clearer settings (\#35). To increase salience, privacy options could be embedded in the installation process. However, many vendors are funded through data collection \cite{Robinson2015}, and therefore might resist alterations. As an accommodation, private settings could be default with alternatives highlighted during installation. Therefore, those who desire functionality can opt-in, while ignorance and apathy would not impede privacy. To monitor the success of such an approach, empirical studies would assess the popularity of different settings. We believe such measures are necessary to reduce the opinion-action disparity.

\section{Conclusions and Further Work}
\label{sec:seven}

In this paper, we explored the Privacy Paradox and the influence of the Internet-of-Things. This is of importance as those who display the disparity might place themselves at risk. Through our 170-person online survey, we discovered that IoT devices are considered significantly less private than non-IoT products. We also found smart devices are regarded as less familiar and (less) usable, with this potentially challenging effective protection. Although the IoT was rated poorly, many who recognised the risks still purchased the products. 

To examine this potential disparity between opinion and action, we conducted contextualised interviews with 40 members of the public. Rather than comparing abstract concepts with practical behaviour, our discussions concerned respondents' devices. We found IoT owners both cared significantly less about their data and were significantly less able to protect it. As supported by our survey results, justifications suggest unfamiliarity and complexity led users to neglect protection.

Directly comparing opinions and actions, we found IoT users were significantly more likely to display the disparity. Seeking to deconstruct the issue, we explored the qualitative rationale of disparity-prone users. Social norms, lack of awareness and short-term necessity were all cited as factors. We concluded by proposing mitigative measures, including IoT awareness campaigns and graphical privacy policies. With a third of our interviewees prone to the disparity, we believe further work is required to mitigate the Privacy Paradox.

We accept our current research possesses several limitations. Our surveys and interviews capture an educated demographic, with a large number of Master's graduates. Although privacy research is often conducted with college-age students, further work will extend these studies with broader demographics. With even these individuals neglecting their data, protection might be rarer for less-educated users. As we phrased our rationale queries in terms of other people, this might have biased responses. While 77\% of disparity-prone respondents referred to themselves, participants might state what they consider to be common replies. In future work, we will use a range of scenarios to dissect why decisions are made.

As mentioned, there is no strict dichotomy between IoT and non-IoT products. However, to explore the influence of smart devices, we selected examples of archetypal products. Future work would extend the range of devices and consider technologies, such as mobile phones, nearer the intersection.

Devices within categories are also diverse, with a Mac desktop differing from a Windows computer. Similar products might offer different privacy settings and collect different pieces of data. By contextualising discussions, we sought to compare each device's concern with its usage. Through identifying disparities at an individual level, we looked to minimise the effect of product diversity. Future work could offer a stricter control by comparing devices from the same vendor. Finally, surveys and interviews are inherently prone to response biases. Through disguising privacy and requesting non-normative opinions, we hope to have minimised their influence. In future work we wish to explore behaviour empirically, comparing actions across a broad range of technologies.

\bibliographystyle{IEEEtran}
\bibliography{bib}

\end{document}